\documentclass{article}

\usepackage{arxiv}

\usepackage[utf8]{inputenc}
\usepackage[T1]{fontenc}

\usepackage[colorlinks=true, linkcolor=blue, citecolor=blue, urlcolor=blue]{hyperref}
\PassOptionsToPackage{hyphens}{url}
\usepackage{url}
\usepackage{booktabs}
\usepackage{amsfonts}
\usepackage{nicefrac}
\usepackage{microtype}
\usepackage{graphicx}
\usepackage{natbib}
\usepackage{doi}

\usepackage{array}
\usepackage{tabularx}
\usepackage{multirow}
\usepackage{colortbl}
\usepackage{xcolor}

\usepackage{amsmath}
\usepackage{amssymb}
\usepackage{textcomp}

\usepackage{makecell}

\usepackage{enumitem}
\usepackage{changepage}
\usepackage{ifthen}
\usepackage{float}

\title{Multi-Agent Firewall Architecture for Privacy Protection of Sensitive Data in Interactions with Language Models}

\author{ {\hspace{1mm}Hugo Garc{\'\i}a Cuesta}\\
	Universidad Carlos III de Madrid\\
	Madrid, Spain \\
\texttt{100428954@alumnos.uc3m.es} \\
	\And
	\href{https://orcid.org/0009-0003-1397-6301}{\includegraphics[scale=0.06]{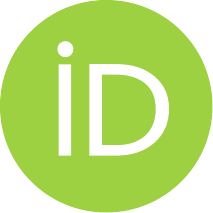}\hspace{1mm}Pablo Mateo Torrej\'on} \\
	Universidad Carlos III de Madrid\\
	Madrid, Spain  \\
	\texttt{pmateo@pa.uc3m.es} \\
    \And
	\href{https://orcid.org/0000-0002-2220-0594}{\includegraphics[scale=0.06]{orcid.pdf}\hspace{1mm}Alfonso S\'anchez-Maci\'an} \\
	Universidad Carlos III de Madrid\\
	Madrid, Spain  \\
	\texttt{alfonsan@it.uc3m.es} \\
}

\renewcommand{\headeright}{}
\renewcommand{\undertitle}{}
\renewcommand{\shorttitle}{Multi-Agent FW Architecture For Protection Of Sensitive Data In Interactions With LLMs
}

\hypersetup{
pdftitle={Multi-Agent Firewall Architecture for Privacy Protection of Sensitive Data in Interactions with Language Models},
pdfsubject={cs.CR, cs.AI},
pdfauthor={Hugo Garc{\'\i}a Cuesta, Pablo Mateo Torrej\'on, Alfonso S\'anchez-Maci\'an},
pdfkeywords={Large Language Models, Prompt Security, Data Leakage Prevention, Named Entity Recognition, Privacy-Preserving Detection, Local Inference},
}

\begin{document}
\maketitle

\begin{abstract}
	While Large Language Models (LLMs) have become essential productivity tools, their integration into workflows without adequate safeguards creates significant risks. This paper proposes an open-source, privacy-focused, user-facing firewall designed to secure both web-based and programmatic LLM interactions. The architecture combines a browser extension and a proxy for total traffic interception across both HTTP(S) and WebSocket communications. At its core, a flexible multi-agent pipeline delivers data leakage prevention through a hybrid approach combining deterministic detectors with LLM-driven semantic analysis, proprietary code leakage prevention, and extensible components designed for future security enhancements such as prompt injection evasion. The framework's layered architecture enables deployment across heterogeneous environments, allowing organizations to balance computational cost, detection depth and latency. Evaluation results demonstrate it achieves F1 scores of up to 94.93\% on optimal configurations.
\end{abstract}


\section{Introduction}
\label{sec:introduction}
Generative Artificial Intelligence has fundamentally shifted how we interact with information, offering a powerful shortcut for complex tasks that previously required hours of human effort. However, this huge leap in capability has outpaced our ability to secure it. As individuals and institutions race to balance these tools, a critical study field has emerged between the immediate appeal of high speed output and the necessity of protecting private information.

While some businesses ban LLMs entirely and others allow unrestricted access, many are stuck in a difficult middle ground. Some attempt to bridge this gap by deploying their own local AI models. However, these internal systems often lack the intelligence of industry leading tools, forcing a choice between cutting-edge power and total data privacy.

Employee behavior remains a critical vector in corporate data security, as professional workflows often prioritize immediate utility and efficiency over security guidelines. This productivity bias is particularly evident among power users who, despite awareness of systemic risks, frequently bypass traditional safeguards to maintain operational speed. Not only do manual policies and static warnings fail to discourage the input of sensitive data, but they also frequently devolve into a source of alert fatigue~\cite{tariq_2025_alert_fatigue}; consequently, human behavior constitutes one of the biggest risks for company security~\cite{hu_2024_llm_compliance}.

This vulnerability is not confined to the corporate perimeter, it extends into the personal domain. Individual users increasingly rely on LLMs for financial planning, legal advice, and health inquiries, often unaware that they are feeding sensitive data into public models. Unlike organizations, which may have dedicated security teams, the average user lacks accessible tools to audit their own AI interactions. This is especially relevant in domestic environments, where a single protective layer could help reduce exposure for all household members, from children using AI for schoolwork to adults handling personal or financial matters.

While existing security strategies are robust for legacy environments, LLMs introduce new challenges. Current approaches each present distinct trade-offs: they either require application modifications, shift trust boundaries to external vendors or compromise on model capability. This work proposes a modular, local-first solution allowing organizations to tailor their security posture depending on operational needs. The following points highlight the core qualities and architectural strengths that define this framework:

\begin{itemize}[leftmargin=*, nosep]
  \item \textbf{Dual Layered Interception}: A hybrid mechanism combining browser-level monitoring for web interfaces with a transparent Man in the Middle (MiTM) proxy for API traffic, including HTTP(S) and WebSocket Secure (WSS) provider channels.
  \item \textbf{Browser Platform Adapters}: The browser extension features an extensible registry pattern allowing rapid integration of support for emerging LLM web interfaces by defining a new platform and targeting specific Document Object Model (DOM) nodes.
  \item \textbf{Multimodal Analysis}: Supports the inspection of both structured text prompts and file uploads to identify sensitive data across a wide variety of formats. The architecture implements a tiered strategy that combines the low-latency of local Optical Character Recognition (OCR) engines for standard document scanning with the semantic reasoning of Vision-Language Models (VLMs) to parse complex screenshots and unstructured visual data.
  \item \textbf{Deeply Customizable Detection Engine}: The core detection logic comes bundled with a predefined set of Personal Identifiable Information (PII) fields alongside an efficient graph based security pipeline. Both the field definitions and the pipeline stages are fully modifiable to meet specific organizational requirements.
  \item \textbf{Code Leakage Prevention}: Custom node that indexes public or private Git repositories and utilizes fuzzy string matching algorithms to detect code leaks, even when proprietary snippets are modified or embedded within natural language queries.
  \item \textbf{Granular Risk Enforcement}: A customizable policy engine that assigns different risk levels to identified threats, enabling different responses such as issuing advisory warnings or blocking requests.
  \item \textbf{Privacy Agentic Workflow}: Beyond its local focused intent, the framework allows dynamic field sanitization across all pipeline agents. This establishes a protective layer for interacting with optional third party detection modules while simultaneously generating a safe output for the end system.
  \item \textbf{Provider Agnostic Inference Backend}: LLM detector agents are fully decoupled from specific providers, enabling organizations to seamlessly switch between local open weight models, proprietary APIs or trusted cloud providers without architectural modifications.
\end{itemize}

The remainder of this paper is organized as follows: Section~\ref{sec:related-work} reviews related work in the field and changes over the time. Section~\ref{sec:system-architecture} details the system architecture, describing each of the packages that constitute the framework. Section~\ref{sec:multiagent-pipeline} provides an in-depth analysis of the multi-agent security pipeline. Finally, Section~\ref{sec:evaluation} evaluates the system's performance over a set of performed tests, followed by conclusions and future work in Section~\ref{sec:conclusions}.

\section{Related Work}
\label{sec:related-work}
Data Loss Prevention (DLP) emerged as a regulatory compliance and intellectual property protection discipline, addressing requirements imposed by frameworks such as GDPR~\cite{gdpr_2016}. Early enterprise solutions from Varonis Systems~\cite{varonis_2005} and Symantec~\cite{symantec_2007} traditionally relied on deterministic pattern matching, fingerprinting, and statistical classifiers to monitor network traffic. Over the years, these providers have progressively integrated semantic analysis and machine learning models to improve document classification and reduce false positives in unstructured data environments.

The integration of Large Language Models (LLMs) pushes beyond this defensive barrier by forcing detection to reach a higher level of semantic understanding. Unlike traditional data exfiltration, LLM interactions involve context rich conversational inputs where proprietary logic can be subtly paraphrased or embedded within long or confusing queries. The OWASP Top 10 for LLM Applications 2025~\cite{owasp_llm_2025} formalizes this gap through Sensitive Information Disclosure (LLM02), which encompasses sensitive data disclosure in user to LLM interactions. While the current implementation focuses on preventing users from inadvertently sending sensitive data to LLMs, the modular architecture is designed for extensibility to future address LLM to user disclosure risks through monitoring of assistant responses.

Modern libraries such as Microsoft Presidio~\cite{microsoft_presidio_2025} achieve high detection rates of structured PII through refined regex and semantic-driven machine learning models like Named Entity Recognition (NER) via SpaCy~\cite{spacy_2020}. While effective for identifying specific tokens without relying on deterministic detectors, a potential expansion of detected items would require retraining of the NER model. Not only does this require significant computational overhead and labeled datasets for every new category, but these systems may also lack the profound semantic reasoning necessary to evaluate the intent of specific prompts. While they can flag isolated entities, they remain oblivious to the risk of contextual exfiltration, where the danger lies not in a single forbidden token, but in the aggregate meaning of an entire message destined for third party LLMs.

Most recent solutions addressing LLM data leakage cluster into four architectural paradigms, each presenting distinct limitations.

The first group are those that provide programmable safety parsers through System Development Kit (SDK) integration. These frameworks enable developers to inject custom scanning logic into application workflows, offering fine-grained control over prompt validation and response filtering. However, their effectiveness is constrained by an interception gap as they require source code modification. Consequently, they cannot protect end users interacting directly with unmanaged web and application interfaces. Great examples are LLM Guard~\cite{llm_guard_2025} and NVIDIA NeMo Guardrails~\cite{nvidia_nemo_2025}, which provide configurable toolkits that allow appending both input and output scanners to user prompts and LLM responses respectively.

\begin{table*}[!t]
  \centering
  \caption{Comparison of representative approaches for sensitive data leakage prevention in LLM interactions.}
  \label{tab:related-work-comparison-matrix}
  \resizebox{\textwidth}{!}{
    \renewcommand{\arraystretch}{2}
    \begin{tabular}{lcccc}
      \toprule
      \textbf{Paradigm \& Representative systems} &
      \makecell[c]{\textbf{Transparent} \\ \textbf{Interception}} & 
      \makecell[c]{\textbf{Semantic} \\ \textbf{Reasoning}} & 
      \makecell[c]{\textbf{Local} \\ \textbf{Analysis}} & 
      \makecell[c]{\textbf{Multimodal} \\ \textbf{Analysis}} \\
      \midrule
      \makecell[l]{\textbf{SDK-integrated guardrails} \\ (LLM Guard; NeMo Guardrails)} &
      No (hooks API calls) & Yes & No & No \\
      \makecell[l]{\textbf{Cloud inspection services} \\ (Lakera Guard; Nightfall AI; Azure AI Content Safety)} &
      Yes (web + API) & Yes & No & Depends \\
      \multicolumn{2}{l}{\makecell[l]{\textbf{Private enterprise local deployment} \\ (Self-hosted open-weight LLM stack)}} & 
      \multicolumn{3}{l}{Doesn't apply. Data does not leave organization boundaries.} \\
      \makecell[l]{\textbf{Local Privacy-Preserving Prompt Assistant} \\ (Zhu et al. 2024)} &
      Yes (web only) & No & Partial (sanitization) & No \\
      \makecell[l]{\textbf{Browser-only interception prototype} \\ (de la Riva, 2025)} &
      Yes (web only) & Yes & No & Partial (PDF only) \\
      \makecell[l]{\textbf{Proposed framework} \\ (Minos Verdict Mesh)} &
      Yes (web + API) & Yes & Yes (sanitization and local analysis) & Yes (PDF, images, code) \\
      \bottomrule
    \end{tabular}
  }
\end{table*}

A different category of frameworks has surged in the recent years as centralized inspection services. Examples of these are Lakera Guard~\cite{lakera_2025}, Nightfall AI~\cite{nightfall_2025}, or Azure AI Content Safety~\cite{azure_content_safety_2025}, which function as external validation layers. Prompts are routed through these services for adversarial prompt injection and sensitive content analysis before reaching the target LLM provider. While effective and usable over third party web chatbots, this architecture creates a privacy paradox as validating the safety of a prompt requires sending its contents to an online hosted vendor, shifting the trust boundary from one company to another.

A third architectural approach, adopted primarily by bigger enterprises, involves completely severing access to third party inference providers in favor of deploying a private LLM infrastructure. While this strategy effectively eliminates data exfiltration risks, it introduces a persistent intelligence gap: organizations must rely exclusively on open weight models, which consistently fall behind proprietary frontier systems.

A fourth approach, mainly emerging from academic research, addresses privacy at the client-side through specialized techniques. The Local Privacy-Preserving Prompt Assistant (LPPA)~\cite{zhu_2024_lppa} operates as a preprocessing tool, assisting users in manually modifying prompts to remove sensitive keywords before submission, focusing on syntactic obfuscation rather than semantic leakage detection. However, LPPA lacks automated semantic detection and relies on an isolated deployment per device, making it unsuitable for organizational scenarios where consistent policy enforcement is required.

The proposed system builds upon the work of de la Riva~\cite{de_la_riva_2025}, who established the viability of LLMs as semantic detectors for PII classification. We advance this foundation by transitioning from a standalone proof-of-concept detector to a comprehensive system featuring transparent interception, multi-layered detection and deep customization capabilities. To address the limitations of prior work, we implement a local-first architecture deployable within corporate infrastructure, home networks, or as a standalone instance on a single device. This design preserves data sovereignty and eliminates external trust dependencies by keeping all detection logic within controlled environments by default. While single-device deployments are constrained by the host's immediate compute power, utilizing a centralized local server allows for more robust detection models, effectively offloading heavy processing from end devices.

This architecture specifically resolves three critical gaps identified across existing approaches through a unified technical strategy. First, it addresses the interception gap via dual-layer traffic capture (detailed in Section~\ref{sec:system-architecture}), enabling protection without requiring application modifications or workflow disruptions. Second, by performing all inspection locally, the system removes the privacy paradox and avoids the trust-boundary shift inherent in cloud-based inspection services. Finally, the design mitigates the intelligence bottleneck through a tiered escalation strategy where deterministic and basic semantic detectors provide low-latency analysis, while complex cases are escalated to local LLM agents for contextual reasoning. This tiered approach further enables advanced features such as multimodal file analysis and proprietary code similarity detection against private repositories.

Data leakage prevention in interactions with LLMs particularly intersects with adversarial attack mitigation. Prompt Injection (LLM01 in OWASP's taxonomy~\cite{owasp_llm_2025}) represents a distinct but related threat where malicious actors manipulate model behavior through crafted inputs. Although the current implementation focuses on data leakage, the modular architecture is designed for extensibility, allowing additional security nodes such as adversarial prompt detection to be integrated without internal refactor of the existing code, positioning the framework as a LLM security platform that could be adapted to new cases that could surge within the following years.

\section{System Architecture}
\label{sec:system-architecture}
The architecture is designed to be extensible and modifiable, decoupling the interception layer (sensors) from the core analysis engine to balance modularity with deployability (see Fig.~\ref{fig:system-architecture-diagram}). Modern LLM usage operates through two dominant channels that each require distinct interception mechanisms: browser webpages for direct human use and programmatic APIs for software integration and agentic use. To support both interaction patterns and evolving protocols like WebSocket-based real-time APIs, the system combines a browser extension with a transparent MiTM proxy covering HTTP(S) and WSS traffic.

\subsection{Deployment Model}
The architecture supports two practical deployment modes depending on the hardware, structure or organization capabilities. In the standalone model, a single user runs the backend and proxy locally while using the browser extension on the same workstation, creating a self-contained privacy layer for both web and API-based LLM interactions. In the centralized model, the backend is hosted on an accessible local server where multiple clients connect to it through the sensors, allowing the same detection logic and policy configuration to be shared across several users or devices. In this second scenario, the implementation supports additional authentication features for security reasons which will be further discussed on following sections.

This flexibility follows from the separation between interception and analysis layers. The sensors remain lightweight and independent, while the backend concentrates the computationally expensive detection logic and policy enforcement. As a result, the same design can operate on commodity personal hardware or as a shared service within a local network.

\subsection{Browser Extension}
The browser extension focuses on intercepting interactions with LLMs in provider websites. It serves a key role as most non power users interact with Generative AI models this way. By having an extension that lives on the user's browser, we ensure traffic from these specific sites can be intercepted.

The extension is compatible with Chromium based browsers which enables seamless deployment across popular platforms like Google Chrome, Microsoft Edge, and Brave, ensuring a broad user reach. By adhering to the recent Manifest V3 standard~\cite{manifest_v3_2025}, the architecture follows enhanced security protocols and efficient resource management, ensuring long-term compatibility with the latest browser privacy policies and performance optimizations.

As discussed in Section~\ref{sec:related-work}, this project builds upon the framework established by de la Riva~\cite{de_la_riva_2025}, who developed an early version of the browser extension as the exclusive interception layer for capturing both textual prompts and PDF uploads within the ChatGPT ecosystem.

\begin{figure*}[t]
  \centering
  \includegraphics[pagebox=trimbox,width=0.95\textwidth,height=0.82\textheight,keepaspectratio]{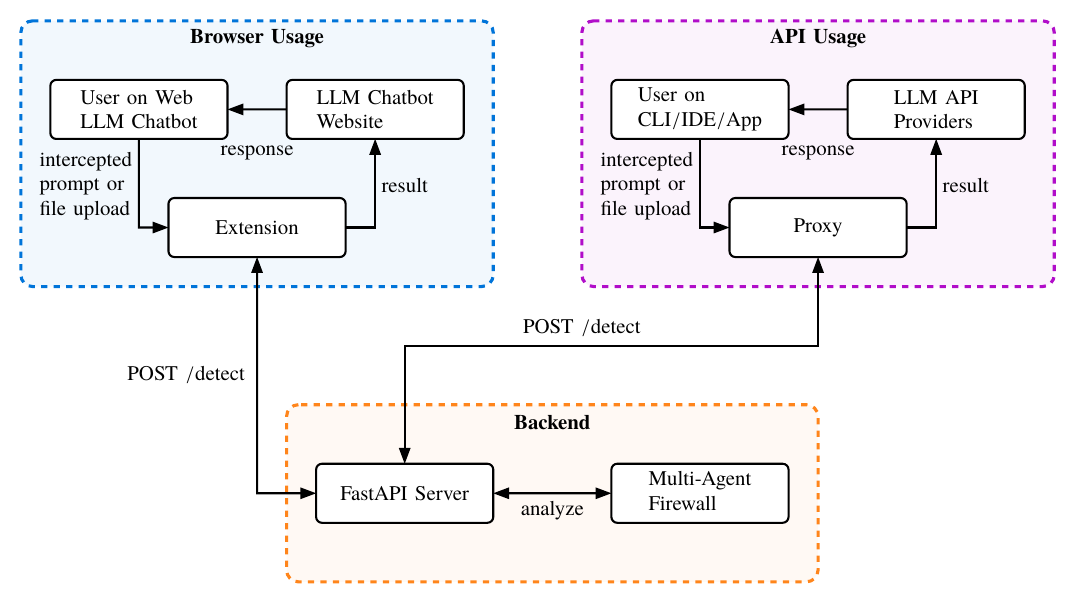}
  \caption{Diagram representation of the system's architecture.}
  \label{fig:system-architecture-diagram}
\end{figure*}

The increasing wave of LLM providers during recent months has demonstrated the need for a way to expand these interception capabilities to different platforms. The biggest obstacle relies on the lack of standardization across web-based LLM interfaces. Provider websites utilize heterogeneous DOM structures and obfuscated CSS classes. To address this, we introduce a modular abstraction layer designed to decouple the interception engine from site-specific frontend implementations, allowing the extension of sensor capabilities to other websites without requiring a complete refactor of the code. The implementation employs an adapter pattern where each supported platform is represented by a class extending a \texttt{BasePlatform} interface. This interface standardizes operations such as identifying the text input composer, locating the send button, extracting and modifying message content, and determining where to inject the risk assessment panel within each platform's unique DOM structure.

Furthermore, the capabilities of the system have been expanded to enable analysis across a wide range of common file formats including image data. This enhancement is facilitated by the integration of OCR engines and VLMs. At the time of evaluation, the system supports PDF, common image formats, and text/code files. Although the file policy is extensible or can be restricted through configuration, adding new binary document formats, such as \texttt{.docx}, may require additional parsing dependencies integration.

A comprehensive technical evaluation of these components is provided in Section~\ref{sec:multiagent-pipeline}. Crucially, the interception logic has been relocated to the pre-upload phase, ensuring that file analysis occurs before data reaches the web platform, preventing the disclosure of sensitive information.

To balance security enforcement with user autonomy, the extension implements a Human-in-the-Loop (HITL) intervention system. Upon detection of sensitive data, the system's response is determined by the risk assessment returned from the security pipeline: high-risk scenarios trigger a blocking modal that prevents message transmission until explicit user action is taken, whereas lower-risk situations present a transient warning notification while permitting prompt transmission. The threshold governing this blocking behavior is configurable through the multiagent firewall package parameters, enabling organizations to balance between security strictness and operational friction according to their risk tolerance.

The assessment panel displays and highlights detected sensitive fields categorized by risk level with data source attribution. On a blocking decision, it allows the user to either follow the advice and cancel the transmission entirely, send an automatically sanitized version where sensitive values are replaced with typed placeholders, or override the security recommendation and proceed with the original prompt. This graduated response mechanism ensures that critical data leaks are gated by human judgment whilst minimizing excessive interruptions.

During the analysis phase, the extension displays a toast notification indicating analysis status whilst simultaneously disabling the send button to prevent premature submission. Upon completion, the notification updates to show the analysis duration, providing transparency into the system's performance. If the backend becomes unreachable during analysis, the extension follows a fail-closed policy and prevents prompt submission rather than allowing an unchecked request to proceed. When the backend is deployed as a shared service, the extension can also attach a bearer token to each request, allowing the same frontend artifact to operate in a secure centrally administered environment.

By design, the extension intentionally maintains a reduced configuration surface, limited to the backend API endpoint, an optional backend authentication token, and the minimum blockage level. This reflects the system's centralized architecture where multiple extension instances connect to a single multiagent-firewall backend that governs all detection logic, enabling centralized management by administrators. The minimum blockage level option is intentionally managed from the sensors side to allow deploying different versions based on technical knowledge and usage expectations from the end users.

\subsection{Proxy}
As discussed in Section~\ref{sec:system-architecture}, the architecture addresses API-based LLM interactions through a transparent MiTM proxy built using the \texttt{mitmproxy} Python package~\cite{mitmproxy_2024}. By configuring system-wide proxy settings via environment variables, outbound API traffic can be routed through this inspection layer, ensuring no programmatic LLM usage bypasses security controls. This includes both classic request-response exchanges over HTTP(S) and persistent WSS sessions used by real-time inference APIs.

The proxy implements selective interception through configurable host and path filtering, analyzing only traffic destined for known LLM providers rather than deep inspecting all network traffic. This selective approach minimizes performance overhead by excluding irrelevant traffic while maintaining extensibility. The library generates a self-signed Certificate Authority (CA) certificate that must be installed in the system's trust store, enabling the proxy to decrypt Transport Layer Security (TLS) traffic.

Text extraction follows a standardized interface across all supported providers. It filters messages by their role field to isolate user-originated content, a convention uniformly adopted by major inference providers. However, image attachment formats lack this standardization. Each provider implements multimodal content differently: OpenAI and Groq embed base64 images as data URLs within the message content array; Anthropic Claude uses a distinct source structure with explicit media type metadata; GitHub Copilot places images in a separate attachments field; and Google Gemini uses an entirely different message structure with ``inline\_data'' encoding. The proxy accommodates this fragmentation through provider-specific extraction logic for each format.

Unlike the browser extension's Human-in-the-Loop workflow, the proxy operates in a fully automated enforcement mode. If the backend's multi-agent pipeline returns a blocking verdict, HTTP requests are rejected with an HTTP 403 response and WebSocket messages are dropped before upstream delivery. If analysis cannot be completed due to malformed payloads or backend unavailability, the proxy follows a fail-closed strategy similar to the extension, rejecting or dropping intercepted content rather than forwarding unchecked data.

For deployments beyond localhost, the proxy exposes two independent authentication boundaries. Communication with the backend may be protected through a shared bearer token attached to every request. On the other hand, client access to the proxy itself may be restricted using mitmproxy's native mechanism backed by an Apache-style htpasswd file, which enables listing multiple permitted users. This approach allows protecting both the sensor to backend channel and the client to proxy ingress path.

Configuration is managed through environment variables supporting key parameters: the backend detection service URL, an optional backend authentication token, the proxy's network binding address and port, an optional timeout threshold for processing delays, and the optional htpasswd file path for client authentication. The core security controls are defined through three comma-separated whitelists: one specifies which API provider domains undergo inspection, one defines the HTTP endpoint patterns to analyze, and one defines WebSocket endpoint patterns. Similar to the extension module, the proxy package allows overwriting the minimum blockage level for distribution with granular policy enforcement.

\subsection{Backend}
As seen on Fig.~\ref{fig:system-architecture-diagram}, the backend subgraph is composed from both the API server and the multi-agent firewall package which contains the detection logic. This section focuses on the behavior of the API server package as the latter will be deeply discussed on Section~\ref{sec:multiagent-pipeline}.

The presented module arises from the necessity of an intermediate node which acts as a glue between the sensors and the security pipeline. It unifies communication across the different interception layers, providing a standardized RESTful interface and enabling asynchronous request handling via the FastAPI library~\cite{fastapi_2024}. This decoupled design also supports independent scaling by facilitating the deployment of multiple API instances across a distributed network, avoiding a bottleneck on high-density sensor environments.

The API exposes a single detection endpoint that processes text, files, or both simultaneously. The backend package treats every uploaded file as untrusted input and applies a layered validation process before forwarding it to the security pipeline. Uploaded content is size limited during streaming to prevent oversized files from being fully written to disk, then checked against the supported type policy before analysis begins. To reduce spoofing risks, the backend verifies consistency between the declared MIME type and the file signature, ensuring that renamed or disguised files are rejected before processing. Temporary files are stored under randomized names mitigating path traversal, overwrite, and unintended execution risks. After analysis completes, these artifacts are automatically removed so sensitive uploaded content does not remain on disk longer than the strictly necessary.

It additionally implements a Cross Origin Resource Sharing (CORS) policy with an allowlist of permitted origins that affects web requests coming from the extension package. This reduces unsolicited browser access to the detection API from arbitrary web origins. CORS alone is not sufficient for a shared deployment, therefore the backend can additionally require the aforementioned bearer token from both the extension and the proxy. Environment variables are safely loaded from the backend side before invoking the security pipeline alongside a per-run minimum blockage level inherited from the sensors.

\section{The Multi-Agent Security Pipeline}
\label{sec:multiagent-pipeline}
The detection engine represents the core module of the proposed framework. It implements a multi-agent architecture based on a hybrid detection strategy orchestrated via a Directed Acyclic Graph (DAG), enabling conditional escalation from deterministic detectors to computationally expensive semantic reasoning agents alongside additional capabilities such as multimodal analysis and proprietary code leakage prevention.

\begin{figure}[h]
  \centering
  \includegraphics[pagebox=trimbox,width=0.8\columnwidth,height=0.7\textheight,keepaspectratio]{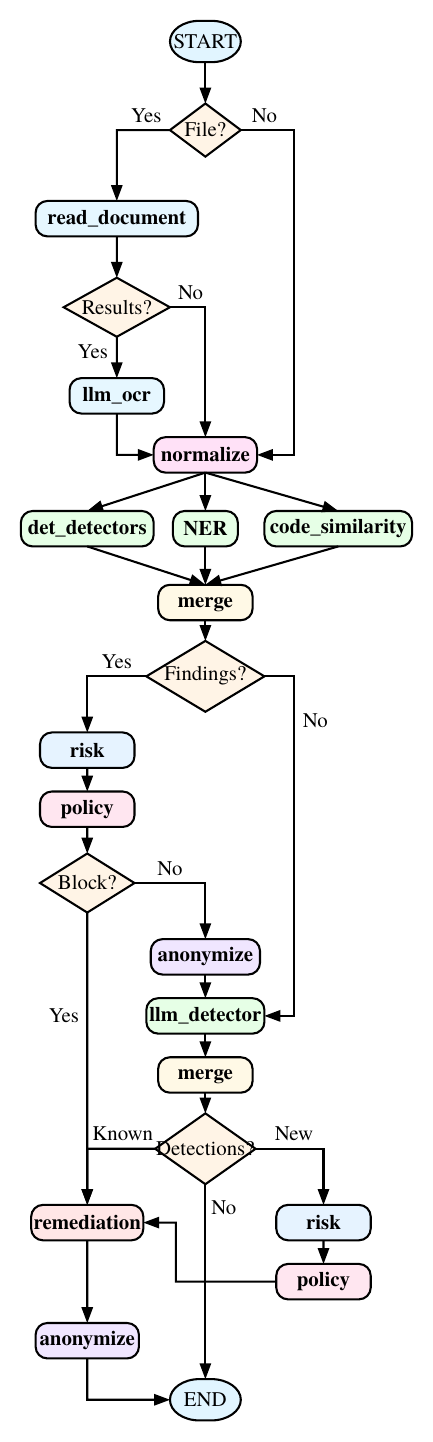}
  \caption{Flowchart of the Detection and Remediation Logic.}
  \label{fig:multiagent-firewall-diagram}
\end{figure}

\subsection{Pipeline Orchestration and Configuration}
The security pipeline is implemented as a state machine using LangGraph~\cite{langgraph_2024}, a framework designed for building stateful workflows with conditional branching and separation of duties. The included detection preset has been tailored to provide time and cost efficient results whilst keeping a high level of detection rates, ensuring it can run on a wide variety of hardware and allowing configuring extras such as file analysis, named entity recognition detectors or proprietary code comparison.

During the design phase, alternative orchestration strategies were evaluated to determine the optimal approach for multi-agent coordination. Specifically, an LLM agent-directed architecture was prototyped, where a language model acted as a dynamic router to select which detection nodes should execute based on runtime state inference. The hypothesis was that agentic decision making could improve adaptability by tailoring inspection paths to input characteristics. However, empirical evaluation revealed significant performance degradation compared to the DAG-based implementation. This LLM orchestrator agent driven variant exhibited higher latency due to the additional inference overhead required for routing decisions, increased cost from auxiliary LLM calls, and reduced determinism in execution paths. Consequently, the static DAG configuration with conditional edges was adopted as the production architecture, providing predictable performance characteristics while maintaining sufficient flexibility through state based branching.

To support a customizable detection approach, the pipeline topology is externalized into a JSON configurable file \texttt{pipeline.json} defining three primitives: nodes, edges, and conditional edges. At initialization this specification is dynamically parsed to build the execution graph. Nodes are resolved through a registry pattern and support custom implementations via Python module path resolution. This architecture enables users to inject proprietary detection logic, duplicate nodes or modify the analysis branching without altering the internal codebase.

Detection behavior is fully configurable through the \texttt{detection.json} specification, which externalizes detection parameters across eight categories: file type extensions and validation constraints, LLM prompt templates for semantic analysis, sensitive field definitions, regular expressions for pattern matching, keyword lists for direct flag of high confidence sensitive values, checksum validators, NER label mappings for entity normalization, and risk scoring rules that classify field severity into high, medium, and low tiers. This externalized configuration enables security teams to adapt detection heuristics to specific security policies or jurisdiction regulations without modifying the detection node implementations.

The modular architecture enables selective deployment of detection capabilities based on operational requirements and resource constraints. Three detection components are available as configurable extras: Named Entity Recognition (NER) for zero-shot semantic entity extraction, file-level analysis for structured document processing, and proprietary code similarity matching for detecting source code exfiltration. Organizations can selectively enable these capabilities during deployment, and the pipeline gracefully degrades when features are disabled. This design philosophy allows resource-constrained environments to run with minimal analysis surface, while bigger infrastructures can activate all layers for maximum coverage.

The pipeline employs a dynamic orchestration layer designed to maximize detection accuracy while minimizing latency and operational costs. Rather than executing an exhaustive linear scan, the system utilizes conditional logic gates to determine the most efficient execution path based on real-time data signals. High-cost components, such as VLMs in file analysis or LLM detectors, are only invoked when first-layer engines provide insufficient confidence or when policy mandates deeper analysis. By pruning the execution DAG when early ``block'' conditions are met, the architecture avoids the ``latency tax'', ensuring that computational resources are strictly reserved for resolving complex semantic ambiguities rather than being applied on every input.

\begin{table}[h]
  \centering
  \caption{Configuration files of the \texttt{multiagent-firewall} package.}
  \label{tab:multiagent-firewall-configuration-files}
  \renewcommand{\arraystretch}{1.5}
  \begin{tabularx}{\columnwidth}{@{}lX@{}}
    \toprule
    \textbf{Configuration file} & \textbf{Purpose} \\
    \midrule
    \texttt{detection.json} & Defines detection rules, supported file types, and risk classification criteria. \\
    \texttt{pipeline.json} & Defines the pipeline graph, including nodes, edges, routers, and execution flow. \\
    \texttt{.env} & Defines runtime parameters such as model providers, OCR, NER, and optional code analysis settings. \\
    \bottomrule
  \end{tabularx}
\end{table}

\subsection{Document Processing}
Following the discussion on Section~\ref{sec:system-architecture}, file uploads are sanitized and validated before reaching the pipeline. However, the system has to convert different file formats into text and provide a unified representation before reaching the detection nodes. PDF documents are processed via pdfplumber~\cite{pdfplumber_2024}. Plain text and code files are read directly with UTF-8 decoding and error handling for malformed encodings. Image files trigger a double tiered OCR strategy designed to balance accuracy and computational cost.

The primary OCR mechanism employs Tesseract~\cite{tesseract_ocr_2025}, an open-source engine capable of extracting text from standard document scans with minimal latency. Tesseract confidence scores are evaluated against a configurable threshold. When Tesseract fails to extract text from image inputs, the conditional router escalates to a Vision Language Model capable of interpreting complex visual layouts, handwritten annotations, and embedded text within graphical contexts. The VLM detector operates through a dedicated and customizable prompt template instructing the model to extract all visible text.

Modern multimodal language models blur the distinction between specialized vision extractors and general-purpose LLMs. Since both the text-based LLM semantic detector and the OCR VLM fallback are abstracted through LiteLLM~\cite{litellm_2024}, organizations can independently configure models for each role or choose to use the same for both. This architectural flexibility enables deploying specialized configurations or a single unified multimodal model for both OCR and contextual analysis, eliminating redundant specialized pipelines while maintaining cost optimization through the two-tier Tesseract fallback strategy.

\subsection{Detection Layer}
The detection strategy implements a two layer architecture that balances throughput and detection depth. The first layer consists of three parallel detector nodes (Deterministic, NER, and code similarity), where the deterministic engine internally combines regex and keyword matching with checksum-aware validation for selected fields. Based on these initial results, the pipeline conditionally escalates to a second layer employing LLM driven semantic analysis, ensuring computationally expensive inference operations are invoked only when required. This parallelization enables concurrent execution of independent detectors while maintaining error isolation.

The findings produced by these layers are then aggregated and transformed into enforceable decisions through the risk scoring and policy mechanisms described in Section~\ref{sec:risk-assessment-and-policy-enforcement}.

\subsubsection{Deterministic Pattern Detectors}
Deterministic detection logic provides high throughput and low latency identification of structured PII through three complementary techniques:

\textbf{Regex Pattern Matching:} The system maintains a customizable library of regular expressions targeting sensitive entities defined in \texttt{detection.json} with optional keyword window validation. This feature requires that regex matches occur within a configurable token distance of keywords to reduce false positives. For example, a 16 digit sequence matching a credit card pattern is only flagged if terms such as ``card'', ``visa'', or ``mastercard'' appear in proximity. This filtering significantly reduces false positives caused by identifiers that are structurally broad and lack a unique signature. External libraries can also be used for regex matching of complex fields such as phone numbers, requiring minimal modification of source code.

\textbf{Checksum Verification as Post-Regex Validation:} For checksum-capable fields, candidates detected by regex are validated before being accepted, filtering structurally plausible but invalid matches. This stricter validation is applied to selected identifiers, reducing false positives while keeping the deterministic execution lightweight.

\textbf{Keyword Matching:} A list of keywords can be defined to search for specific strings of text that should trigger a detection. Unlike pattern-based discovery, this mechanism performs exact string matching to flag high confidence identifiers.

\begin{figure}[h]
  \centering
  \includegraphics[pagebox=trimbox, width=0.3\textwidth]{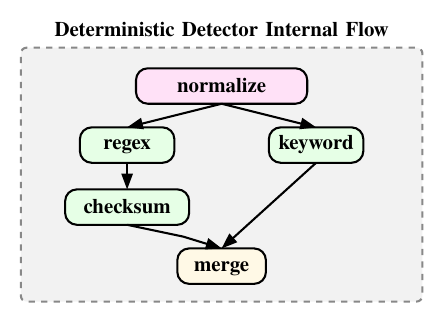}
  \setlength{\abovecaptionskip}{1pt}%
  \caption{Deterministic Detector Internal Flow.}
  \label{fig:pattern-detectors-diagram}
\end{figure}

\subsubsection{Zero-Shot Named Entity Recognition}
Semantic entity extraction is performed via GLiNER~\cite{gliner_2024}, a Named Entity Recognition (NER) model trained on diverse multilingual datasets. Unlike Microsoft Presidio's SpaCy which requires specific training on labeled sets, GLiNER operates in zero-shot mode by accepting arbitrary entity labels at inference time. This capability eliminates the retraining bottleneck when coverage needs to be extended to novel entity categories or domain specific terminology. It additionally allows to set a confidence threshold to balance precision and recall.

Overall, this NER agent provides a basic semantic detection layer that captures context-dependent entities which lack a predictable structural format, serving as a flexible fallback for identifying sensitive information that regular expressions alone fail to resolve.

\subsubsection{Heuristic Fuzzy Code Similarity Matching}
Proprietary source code represents a critical exfiltration risk in AI assisted development workflows. Traditional keyword and NER detectors fail to identify such leaks because code snippets lack the structured patterns characteristic of PII and are often embedded within natural language descriptions or modified through variable renaming and refactoring.

The code similarity detector addresses this threat through fuzzy string matching against indexed Git repositories and file extensions. When analyzing input text, the detector extracts potential code segments through heuristic parsing. Each extracted segment undergoes fuzzy matching against the indexed corpus via the RapidFuzz library~\cite{rapidfuzz_2024}, which implements optimized string similarity metrics including Levenshtein distance.

\subsubsection{LLM-Based Contextual Reasoning}
The final detection layer employs a language model agent to perform semantic reasoning over anonymized prompt contexts, addressing the intelligence gap inherent in deterministic and zero-shot NER approaches.

The LLM detector receives a configurable system prompt that dynamically enumerates the canonical field taxonomy and instructs the model to identify both explicit mentions and contextually inferred sensitive entities. By operating over anonymized inputs (see Section~\ref{sec:sanitization-workflow}), sensitive values are protected while enabling semantic reasoning.

Detected entities are attributed to sources \texttt{llm\_explicit} (directly mentioned) or \texttt{llm\_inferred} (contextually derived), enabling auditability of model reasoning. To prevent circular detection loops, the detector implements filtering logic that discards any findings containing redaction tokens (e.g., \texttt{<<REDACTED:SSN\_1>>}).

Recent industry developments further validate the viability of small, locally deployable models for semantic PII detection. OpenAI's recently released Privacy Filter~\cite{openai_privacy_filter_2026} as a 1.5B parameter open-weight model, demonstrating that frontier-level detection performance is achievable without requiring large-scale inference infrastructure. The mentioned model achieves 96\% F1 on the \texttt{PII-Masking-300k}. This model is an excellent candidate to integrate as an LLM detection node over the discussed architecture, however, its fixed taxonomy of eight categories would underutilize the framework's core design of extensible and configurable detection, preventing organizations from adapting the semantic reasoning layer to custom PII entities.

\subsection{Sanitization-First Workflow}
\label{sec:sanitization-workflow}
A fundamental challenge in deploying LLM-based detection arises from the privacy paradox discussed in Section~\ref{sec:related-work}: the security system requires access to sensitive data to identify threats, yet exposing that data to external models defeats the framework's confidentiality guarantees. The architecture resolves this through a sanitization-first workflow where deterministic and NER findings are redacted before contextual analysis by LLM agents.

The anonymization subsystem operates prior to the two potential exfiltration points: the LLM detector (when relying on an external provider) and the destination LLM endpoint. These nodes replace each detected value with a typed placeholder token following the format \texttt{<<REDACTED:FIELD\_NAME\_N>>}, where \texttt{N} is an incremental counter ensuring unique tokens for multiple instances of the same field type.

This anonymized text is passed to the LLM detector, which performs semantic reasoning over redacted tokens rather than actual values. After detection, a second anonymization phase applies redaction to any newly detected entities, producing fully sanitized output. This approach enables optional third-party LLM integration without exposing sensitive values in clear text, and allows users to manually transmit sanitized variants through the extension interface when appropriate.

\subsection{Risk Assessment and Policy Enforcement}
\label{sec:risk-assessment-and-policy-enforcement}
The policy enforcement subsystem translates detected entities into actionable security decisions through a two-stage process: risk scoring and threshold-based policy application.

Risk scoring implements a weighted classification system defined in \texttt{detection.json} under the \texttt{risk\_fields} schema. Sensitive fields are categorized into three tiers: high risk, medium risk, and low risk. Each tier is assigned a numeric weight and the cumulative risk score is computed by adding the weights across all detected fields. A separate object defines the threshold from where an analysis is globally considered as low, medium or high risk. This scoring mechanism recognizes that multiple low risk entities in combination may constitute a higher aggregate threat than any single field in isolation.

Policy enforcement compares the computed risk level against the \texttt{min\_block\_level} threshold, configurable per-request from each of the sensors. If the globally computed risk exceeds the minimum block level the decision is set to ``block''. If detected fields exist but fall below the threshold, the decision is set to ``warn''. If no fields are detected, the decision is ``allow''.

The remediation agent generates feedback messages appropriate to the enforcement decision. Blocking messages enumerate the detected field types and instruct the user to redact or remove flagged content. Warning messages provide similar field listings with advisory language recommending redaction. These messages are surfaced through the browser extension's modal interface or, in the case of API proxy interception, as structured error responses with HTTP 403 status codes.

This extensibility model positions the framework as a mesh of security agents rather than a fixed DLP appliance, enabling adaptation to emergent threat vectors such as prompt injection through modular composition rather than monolithic refactoring.

\section{Evaluation \& Benchmarks}
\label{sec:evaluation}
\subsection{Evaluation Scope \& Design}
This evaluation assesses the core multi-agent detection pipeline's ability to identify sensitive data across textual prompts with practical balance between detection quality, latency, and privacy. In contrast with prior work centered on LLM detection over cloud hosted models such as in de la Riva's thesis~\cite{de_la_riva_2025}, the objective is not to maximize raw semantic capability but to determine how far a hybrid pipeline can be pushed when restricted to smaller, locally deployable models.

It excludes several capabilities of the framework such as multimodal analysis, code similarity detection or anonymization fidelity which would be harder to evaluate accurately. It was decided to prioritize sensitive data leakage as it remained the original scope of the thesis, leaving these evolved secondary features to be evaluated either in isolation or in conjunction in future work.

The experiments are executed through the \path{multiagent-firewall} package over the \path{nvidia/Nemotron-PII}~\cite{nvidia_nemotron_pii_2025} test split, restricted to the \texttt{us} locale and capped at 500 cases per run. All evaluated configurations operate over an identical sample selected from a fixed seed, ensuring comparability and reproducibility across configurations. The LLM employed for semantic detection is configured with deterministic inference parameters to prevent model randomness from causing architectural differences. Detailed hardware specifications, software stack versions, and exact LLM parameters are documented in Appendix~\ref{sec:appendix-b}.

To ensure fair scoring against the provided dataset labels, a normalization layer converts annotations from snake\_case to the framework's uppercase taxonomy (e.g., \texttt{credit\_debit\_card} $\rightarrow$ \texttt{CREDIT\_DEBIT\_CARD}). During matching, detected entities are compared using a flexible equivalence rule accepting exact matches, substring matches, and predefined bidirectional aliases for semantically close categories. For example, \texttt{IPV4} counts as \texttt{IP} and \texttt{FIRST\_NAME} as \texttt{PERSON}. This prevents penalizing cases where detectors identify a subset or provide a broader semantic classification that is technically accurate despite nomenclature differences.

\subsection{Detection Performance Analysis}

The ablation table (see Table~\ref{tab:evaluation-ablation-table}) summarizes the results from the tests carried out. Each group is analyzed below to isolate the effect of specific design decisions.

\subsubsection{Group 1: Minimal Approach}
The \texttt{T1} configuration represents the original and minimal thesis approach, combining deterministic detectors with conditional LLM semantic escalation only when deterministic methods require additional reasoning depth. As the minimum blockage level for the tests is set to the most restrictive one (``low'') it means that LLM detection is only triggered if no entities have been found for the current prompt. This theoretically leaves room for entities not found by the deterministic detectors to stay unidentified, however, the behavior of the pipeline would still block the request and therefore achieve the data leakage prevention objective saving additional inference costs.

\texttt{T1} achieves 93.40\% precision with only 47.33\% recall (F1 62.83\%) and extremely low latency (mean 38.33 ms). The high precision reflects that regex-based matching with keyword validation produces few false positives, while the recall gap reveals that deterministic patterns alone miss nearly half of sensitive entities. This establishes the motivation for additional semantic detectors or forced LLM analysis in subsequent configurations.

\subsubsection{Group 2: NER Threshold Sweep}
These configurations introduce zero-shot Named Entity Recognition (NER) through GLiNER~\cite{gliner_2024} at progressively higher confidence thresholds (0.3, 0.5, 0.7 and 0.85 respectively), enabling study of the precision-recall trade-off in semantic detection.

Thresholds 0.3, 0.5, 0.7 deliver comparable performance with F1 scores clustering around 81--82\%, substantially better than \texttt{T1}'s 62.83\%. The three configurations show minimal latency differences of around 800ms each. Even though it significantly increases latency from \texttt{T1}'s settings, it still remains within acceptable bounds for real-time applications. Within this cluster, slight precision-recall shifts occur as threshold increases, but the differences are marginal. Any of these three represents a viable conditional configuration for interactive scenarios.

As the NER threshold increases to 0.85 (\texttt{T5}), precision improves to 92.45\% but recall drops significantly to 58.68\% (F1 71.79\%), representing a different trade-off profile suited only to precision-critical deployments.

Conditional LLM inference is rarely triggered in this group, validating that NER alone identifies enough PII for triggering a blockage decision.

\subsubsection{Group 3: Exhaustive Detection}
These configurations force all detectors to run, directly testing whether additional semantic reasoning compensates for its computational cost. The aim of these tests is to check whether the LLM detectors can be useful to catch sensitive fields that are skipped by the rest of the detectors and if the selection of different models makes an impact. Based on results from group 2, all of the remaining tests will be run with a NER threshold of 0.5.

\texttt{T6} evaluates the Gemma 3 4B model (instruction tuned)~\cite{google_gemma_3_4b_it_2024}, achieving 84.55\% precision, 86.56\% recall, and 85.54\% F1. Relative to prior best results, this represents +8.23 recall points at the cost of more than a second in mean latency ($\sim$2x slower). The $\sim$8 point recall improvement comes with no precision benefit and a substantial latency cost, indicating that conditional LLM execution only for ambiguous cases remains more practical than forced exhaustive detection.

\texttt{T7} evaluates Ministral 3 8B~\cite{mistralai_ministral_3_8b_instruct_2512_2025}, reaching 85.10\% precision, 87.78\% recall and 86.42\% F1 with 1.258 seconds mean latency. Compared with \texttt{T6}, it improves both recall and F1 with a lower latency profile, suggesting a better balance for environments that can handle inference from a bigger model.

\texttt{T8} replaces Gemma with Llama 3.1 8B~\cite{meta_llama_3_1_8b_2024}. Compared with \texttt{T7}, it shows lower recall and lower F1 with almost identical precision, while latency increases sharply to 5.96 seconds. This reinforces that larger models do not necessarily translate into better detection quality and that they should be tested to determine if the performance trade-offs align with operational constraints.

\subsubsection{Group 4: LLM-Only and Advanced Models}
This final group explores the upper bounds of performance by disabling NER and relying solely on forced LLM driven detections.

\texttt{T9} evaluates Gemma 3 4B again, achieving 83.28\% F1 with a mean latency of 0.8 seconds. Compared to Group 3 \texttt{T6}, this shows that for smaller models the hybrid NER+LLM approach provides a 2-point F1 advantage even when recall decreases by almost 10 points. There is, however, a latency improvement that could be considered for environments prioritizing responsiveness.

\texttt{T10} employs a version of the Gemma 3 4B model fine-tuned over the training subset of the \texttt{Nemotron-PII} dataset using Unsloth~\cite{unsloth_2024}. This configuration achieves the highest performance across all tests, reaching 94.93\% F1 with balanced precision (94.95\%) and recall (94.92\%). While the mean latency is higher at more than 5 seconds, the breakdown shows a median of only 0.66 seconds, indicating that high latency is concentrated in outliers. This result demonstrates that fine-tuning can significantly enhance the precision-recall balance of lightweight models, enabling them to rival much larger architectures.

\texttt{T11} tests the recently released Gemma 4 E4B model~\cite{google_gemma_4_e4b_it_2026}, achieving 89.43\% F1. Notably, it delivers exceptional precision (95.21\%) and the lowest latency among semantic-heavy configurations (0.53 seconds). The latency difference could come from updates on the model serving performed by vLLM, as it had to be updated to a newer version for it to successfully serve the model. The improvements compared with its predecessor support the thesis that as smaller models continue to gain intelligence, they will reach sufficient capabilities for data leakage prevention tasks while maintaining the efficiency required for real-time deployment.

\begin{table*}[h]
  \centering
  \caption{Benchmark results across all evaluated configurations.}
  \label{tab:evaluation-ablation-table}
  \resizebox{\textwidth}{!}{
    \begin{tabular}{lccccccc}
      \toprule
      \textbf{Config} & \textbf{NER} & \textbf{LLM} & \textbf{LLM model} & \textbf{Prec.} & \textbf{Rec.} & \textbf{F1} & \textbf{Lat. (s)} \\
      & \textbf{thresh.} & \textbf{exec.} & & & & & \\
      \midrule
      \multicolumn{8}{c}{\textit{Group 1}} \\
      \texttt{T1-DLP-COND-LLM} & N/A & Conditional & \texttt{gemma3:4b} & 93.40\% & 47.33\% & 62.83\% & 0.038 \\
      \midrule
      \multicolumn{8}{c}{\textit{Group 2}} \\
      \texttt{T2-DLP-NER-30} & 0.3 & Conditional & \texttt{gemma3:4b} & 85.56\% & 78.10\% & 81.66\% & 0.834 \\
      \texttt{T3-DLP-NER-50} & 0.5 & Conditional & \texttt{gemma3:4b} & 85.56\% & 78.10\% & 81.66\% & 0.849 \\
      \texttt{T4-DLP-NER-70} & 0.7 & Conditional & \texttt{gemma3:4b} & 90.00\% & 69.42\% & 78.38\% & 0.845 \\
      \texttt{T5-DLP-NER-85} & 0.85 & Conditional & \texttt{gemma3:4b} & 92.45\% & 58.68\% & 71.79\% & 0.868 \\
      \midrule
      \multicolumn{8}{c}{\textit{Group 3}} \\
      \texttt{T6-FORCE-GEMMA} & 0.5 & Forced & \texttt{gemma3:4b} & 84.55\% & 86.56\% & 85.54\% & 2.175 \\
      \texttt{T7-FORCE-MINISTRAL} & 0.5 & Forced & \texttt{ministral3:8b} & 85.10\% & 87.78\% & 86.42\% & 1.258 \\
      \texttt{T8-FORCE-LLAMA} & 0.5 & Forced & \texttt{llama3.1:8b} & 85.16\% & 86.01\% & 85.59\% & 5.959 \\
      \midrule
      \multicolumn{8}{c}{\textit{Group 4}} \\
      \texttt{T9-ONLY-GEMMA} & N/A & Forced & \texttt{gemma3:4b} & 90.94\% & 76.82\% & 83.28\% & 0.807 \\
      \texttt{T10-ONLY-GEMMA-FT} & N/A & Forced & \texttt{gemma3-ft} & 94.95\% & 94.92\% & 94.93\% & 5.419 \\
      \texttt{T11-ONLY-GEMMA-4} & N/A & Forced & \texttt{gemma4:e4b} & 95.21\% & 84.31\% & 89.43\% & 0.528 \\
      \bottomrule
    \end{tabular}
  }
\end{table*}

\begin{figure}[h]
  \centering
  \includegraphics[pagebox=trimbox, width=0.5\textwidth]{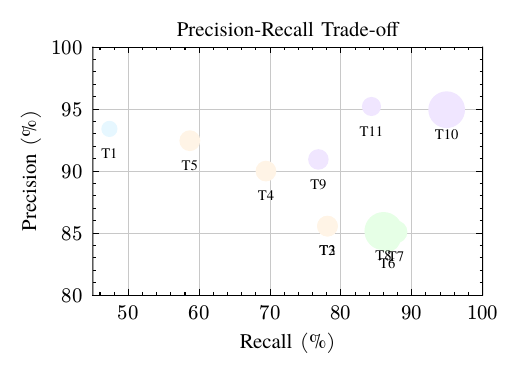}
  \setlength{\abovecaptionskip}{1pt}%
  \caption{Precision vs recall trade-off, with bubble size proportional to latency.}
  \label{fig:performance-tradeoff-chart}
\end{figure}

\subsection{Detector Contribution Analysis}

Table~\ref{tab:tp-fp-breakdown-table} displays findings contribution per source, revealing how the layered architecture behaves internally and allowing the extraction of the following key insights.

\begin{table*}[h]
  \centering
  \caption{Per-source true positives and false positives contribution breakdown. Each cell shows TP/FP (FP\%) format, where the percentage represents the false positive rate for that detector.}
  \label{tab:tp-fp-breakdown-table}
  {\scriptsize
  \resizebox{\textwidth}{!}{
    \begin{tabular}{lccccccc}
      \toprule
      \textbf{Config} & \textbf{Reg. Expr.} & \textbf{Checksums} & \textbf{Phone\#} & \textbf{NER} & \textbf{LLM Exp.} & \textbf{LLM Inf.} & \textbf{Other} \\
      \midrule
      \texttt{T1} & 1518/32 (2\%) & 37/0 (0\%) & 218/56 (20\%) & 0/0 (--) & 78/19 (20\%) & 2/3 (60\%) & 1/0 (0\%) \\
      \texttt{T2} & 1518/32 (2\%) & 37/0 (0\%) & 218/56 (20\%) & 2249/461 (17\%) & 1/1 (50\%) & 0/0 (--) & 0/0 (--) \\
      \texttt{T3} & 1518/32 (2\%) & 37/0 (0\%) & 218/56 (20\%) & 2249/461 (17\%) & 1/1 (50\%) & 0/0 (--) & 0/0 (--) \\
      \texttt{T4} & 1518/32 (2\%) & 37/0 (0\%) & 218/56 (20\%) & 1131/191 (14\%) & 6/6 (50\%) & 0/0 (--) & 0/0 (--) \\
      \texttt{T5} & 1518/32 (2\%) & 37/0 (0\%) & 218/56 (20\%) & 436/73 (14\%) & 13/6 (32\%) & 0/0 (--) & 0/0 (--) \\
      \texttt{T6} & 1518/32 (2\%) & 37/0 (0\%) & 218/56 (20\%) & 2249/461 (17\%) & 367/51 (12\%) & 9/10 (53\%) & 4/35 (90\%) \\
      \texttt{T7} & 1518/32 (2\%) & 37/0 (0\%) & 218/56 (20\%) & 2249/461 (17\%) & 396/86 (18\%) & 9/5 (36\%) & 0/0 (--) \\
      \texttt{T8} & 1518/32 (2\%) & 37/0 (0\%) & 218/56 (20\%) & 2249/461 (17\%) & 454/122 (21\%) & 3/5 (63\%) & 0/0 (--) \\
      \texttt{T9} & 1518/32 (2\%) & 37/0 (0\%) & 218/56 (20\%) & 0/0 (--) & 1067/157 (13\%) & 17/20 (54\%) & 4/5 (56\%) \\
      \texttt{T10} & 1518/32 (2\%) & 37/0 (0\%) & 218/56 (20\%) & 0/0 (--) & 1653/78 (5\%) & 0/0 (--) & 0/0 (--) \\
      \texttt{T11} & 1518/32 (2\%) & 37/0 (0\%) & 218/56 (20\%) & 0/0 (--) & 1302/53 (4\%) & 0/1 (100\%) & 0/0 (--) \\
      \bottomrule
    \end{tabular}
  }
  }
\end{table*}

Establishing a deterministic environment over the same seed and number of cases leads the deterministic detectors to be consistent across all runs, yielding $\sim$2\% of false positives. The checksum engine, which acts as a double check for social security numbers, offers perfect detections with no false positives. The Python's port of Google's libphonenumbers library is an example of an external library acting as a deterministic detector over phone number matching, being one of the hardest regex patterns to match due to the high variability of international formats and therefore having a higher false positive rate.

\begin{figure}[h]
  \centering
  \includegraphics[pagebox=trimbox]{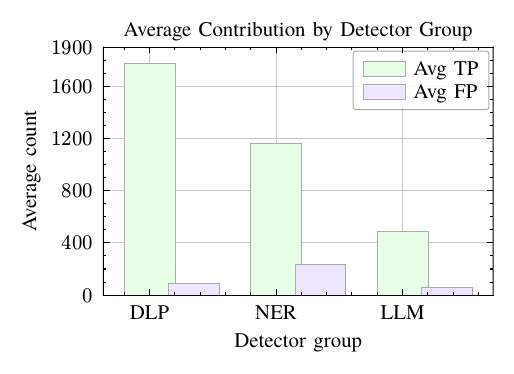}
  \setlength{\abovecaptionskip}{1pt}%
  \caption{Average TP/FP contribution by detector group cross evaluated configurations.}
  \label{fig:detector-contribution-chart}
\end{figure}

The NER model is the primary recall driver when enabled. Conditional configurations with lower thresholds yield up to 2249 TP with 461 FP, demonstrating substantial recall gain at the cost of elevated false positives. Higher thresholds restrict contributions to 436 TP with only 73 FP, validating the precision-recall trade-off. The elevated percentage of false positives in lower-threshold configurations highlights the challenge of zero-shot entity extraction without post-filtering.

Within the exhaustive NER+LLM group (\texttt{T6}--\texttt{T8}), forcing all detectors improves semantic coverage compared with conditional execution, but consistently increases latency and false-positive pressure. The main difference between models is therefore not architectural behavior but operating point: each one trades quality and responsiveness differently. In practice, this group confirms that model selection defines the precision-recall-latency balance if full semantic analysis is enabled.

In environments without NER, the LLM detector becomes the primary and only semantic source. On forced executions it remains a strong alternative for recovering entities that deterministic rules miss, although with a higher false-positive burden in non fine-tuned models. Inferred contributions remain a minority, showing that implicit-signal extraction is possible but still noisy.

The source \texttt{other} was set as a fallback for cases where LLM detectors failed to categorize attribution to either explicit or inferred. This is expected to happen when models hallucinate and fallback prevented the detection from being discarded. This behavior appears in Gemma 3 runs and by inspecting the logs we see traces with extracted values like \texttt{LAST\_NAME = None} or \texttt{POSTCODE = None} which supports the hypothesis. This also explains its high false-positive ratio and motivates rejection of detections attributed to this label.

\section{Conclusion \& Future Work}
\label{sec:conclusions}
This work demonstrates that a local-first, multi-agent architecture can provide practical data leakage prevention for LLM interactions without requiring application code changes or routing sensitive content through external scanning vendors. By combining a browser extension and a transparent MiTM proxy, the system covers both dominant interaction channels in current LLM ecosystems.

At the detection layer, the proposed DAG pipeline validates a hybrid strategy where deterministic and lightweight semantic components provide broad low-latency coverage, while LLM-based contextual reasoning is applied through conditional escalation only when additional semantic depth is justified. This design preserves operational efficiency while keeping extensibility as a core property of the framework.

The sanitization-first workflow further resolves the privacy paradox of semantic detection by allowing advanced reasoning over redacted content, maintaining data sovereignty even when optional third-party model providers are configured. Together, these architectural properties position the artifact as a deployable security layer for both organizational and personal environments where unrestricted LLM access is no longer acceptable.

\subsection{Evaluation Takeaways}
The evaluation performed in Section~\ref{sec:evaluation} establishes clear performance profiles for distinct deployment scenarios:

For environments where latency is the priority, conditional configurations such as \texttt{T2} and \texttt{T3} achieve good results at around 85.56\% precision, 78.10\% recall, and 81.66\% F1 with mean response times of less than a second. These configurations optimize the interactive-use balance, combining deterministic foundation with aggressive NER recall amplification at low confidence thresholds, while constraining conditional LLM escalation to minimal additional detections.

If the priority shifts to PII detection, \texttt{T10} achieves 94.95\% precision, 94.92\% recall, and 94.93\% F1 at the cost of 5419.32 ms mean latency. This configuration demonstrates that locally-deployable fine-tuned models, can match or exceed the semantic reasoning capacity of larger general-purpose systems. The extremely low false-positive rate and high true-positive count indicate that model specialization drives detection quality at this scale.

\subsection{Limitations and Future Work}
Although local inference is a deliberate design decision to maximize privacy and deployment control, it also introduces a quality ceiling in difficult semantic cases when compared with larger hosted frontier models. As AI laboratories keep improving their smaller open weights models, it is expected that the gap between edge-device performance and cloud-scale intelligence will narrow up to the point of contextual sufficiency. Along this line, future iterations should benchmark newly released NER models and LLMs, specifically those fine-tuned for sensitive data leakage prevention, quantifying how these updates shift the precision-recall-latency frontier under an identical evaluation protocol.

Browser-side coverage is also constrained by the current implementation scope, which targets Chromium-based extension APIs and a limited set of provider adapters. This limitation is partially mitigated by the fact that Chromium browsers represent the dominant share of desktop usage (around 87\%~\cite{statcounter_desktop_browser_share_2026}), leaving a potential motivation on expanding the extension capabilities to Gecko or WebKit based browsers.

The current API interception scope focuses on HTTP(S) and WSS channels. This design choice aligns with industry standards, as major LLM providers focus almost exclusively on these as their supported integration protocols as seen on OpenAI's Realtime API~\cite{openai_realtime_api_2026} and Google's Vertex AI Live API~\cite{google_vertex_ai_live_2026}. The proxy stack can evolve with optional modules for gRPC, gRPC-web, and selected WebRTC communications in the case these continue to grow in adoption.

The architecture would also benefit from introducing Human-in-the-Loop controls for API originated traffic, with the caveat of having to deal with the absence of a unified presentation layer, requiring to handle approvals from headless callers like CLI tools, IDEs and backend services. The proxy subsystem should also evolve towards an extensible plugin registry, mirroring the extension adapter pattern, so users can add and maintain payload parsers without modifying internal code.

By utilizing the framework's extensibility capabilities, we could extend monitoring to assistant responses and introduce dedicated prompt injection defenses, ensuring broader coverage under a unified policy engine.

Future benchmarks should explicitly measure document parsing fidelity, code similarity analysis and anonymization capabilities. These three evaluation axes would quantify how extraction and transformation quality propagates to final risk decisions across diverse situations.

\section*{Acknowledgments}
The work has been partially supported by R\&D project PID2022-136684OB-C21 (Fun4Date-Redes) funded by the Spanish Ministry of Science and Innovation MCIN/AEI/10.13039/ 501100011033 and TUCAN6-CM (TEC-2024/COM-460), funded by Comunidad de Madrid (ORDEN 5696/2024).

The authors would like to thank SLICES-Madrid (https://slices-madrid.eu/), the main site of SLICES-Spain, part of the ESFRI SLICES-RI project, for the use of AI Training Research Infrastructure.

\bibliographystyle{unsrt}  
\bibliography{refs}

\clearpage
\appendix
\section{Source Code, Extensibility and Dependencies}
\label{sec:appendix-a}

\subsection{Repository Access}
The complete implementation is available at:
\vspace{0.2cm}
\begin{center}
  \url{https://github.com/xHugo21/minos-verdict-mesh}
\end{center}
\vspace{0.2cm}
The latest commit hash at the time of publishing this paper is:
\begin{center}
  \texttt{4ee6c563a84d24e2494c1437aaefdc947dc00e05}
\end{center}

\subsection{Extensibility Surfaces}
The project exposes multiple extension points intended for long-term evolution:
\begin{itemize}[nosep]
  \item \textbf{Browser platform adapters}: New web providers can be integrated by implementing a \texttt{BasePlatform} compatible adapter and adding platform specific DOM mappings.
  \item \textbf{Pipeline customization}: Custom detector nodes and routing functions can be registered through \path{config/registry.py} and referenced from \path{pipeline.json}.
  \item \textbf{Detection taxonomy and validation logic}: Detection rules can be fully customized from \path{detection.json} via regex patterns, keyword lists, risk mappings, and checksum validator bindings.
  \item \textbf{Library based detector hooks}: Complex field extraction can be integrated through library sentinels in \texttt{detection.json}.
  \item \textbf{Proxy interception formats}: Host and path allowlists for HTTP(S) and WSS interception can be extended to support additional provider protocols without changing the backend API contract.
  \item \textbf{Deployment model composition}: Extension-only, proxy-only, or combined operation can be chosen depending on organizational usage channels.
  \item \textbf{Authentication}: Bearer token authentication for sensor-to-backend communication and client authentication for proxy access can be customized via environment variables, htpasswd files, or extended with organization-specific mechanisms such as OAuth or API key validation.
\end{itemize}
This design enables controlled evolution of the architecture without refactoring the core orchestration engine.

\subsection{Configuration Artifacts}
The core firewall configuration files are presented in Table~\ref{tab:multiagent-firewall-configuration-files} within Section~\ref{sec:multiagent-pipeline}. Together, \texttt{pipeline.json} and \texttt{detection.json} define graph orchestration, detection taxonomy, and risk scoring behavior without requiring code-level changes.

At project level, runtime behavior is additionally controlled through package specific configuration files:
\begin{itemize}[nosep]
  \item \path{multiagent-firewall/.env} for detector model selection and optional analysis modules.
  \item \path{backend/.env} for API server settings and backend authentication.
  \item \path{proxy/.env} for interception scope, timeout, and optional proxy-side client authentication.
  \item \path{extension/src/config.js} for browser sensor endpoint and minimum blocking policy.
\end{itemize}

\subsection{Dependencies and Licensing}
The entire framework is released under an MIT license, extending the permissive licensing of the original work by de la Riva~\cite{de_la_riva_2025}. This choice facilitates broad research adoption and commercial flexibility by minimizing redistribution constraints. The remainder of this section provides a breakdown of the dependencies employed to ensure full transparency regarding their respective legal and distribution requirements.

The system integrates third-party components governed by permissive licenses (MIT, BSD-3-Clause, Apache 2.0). The Software Bill of Materials (SBOM) is available as \texttt{sbom.json} within the source code.

\makeatletter
\setlength{\@fptop}{0pt}
\makeatother
\begin{table}[!ht]
  \centering
  \caption{First-level dependencies and associated licenses by module.}
  \label{tab:project-dependencies}
  {\scriptsize
  \resizebox{0.6\columnwidth}{!}{
    \begin{tabular}{llll}
      \toprule
      \textbf{Module} & \textbf{Dependency} & \textbf{Version} & \textbf{License} \\
      \midrule
      \multirow{3}{*}{\textbf{Proxy}} & httpx & v0.28.1 & BSD-3 \\
      & mitmproxy & v10.3.0 & MIT \\
      & python-dotenv & v1.2.1 & BSD-3 \\
      \midrule
      \multirow{11}{*}{\textbf{Multiagent Firewall}} & langchain-litellm & v0.3.2 & MIT \\
      & langgraph & v0.5.4 & MIT \\
      & phonenumbers & v9.0.22 & Apache 2.0 \\
      & typing-extensions & v4.15.0 & PSF \\
      & gitpython & v3.1.46 & BSD-3 \\
      & rapidfuzz & v3.14.3 & MIT \\
      & pdfplumber & v0.11.8 & MIT \\
      & pillow & v12.0.0 & HPND \\
      & pytesseract & v0.3.13 & Apache 2.0 \\
      & filetype & v1.2.0 & MIT \\
      & gliner & v0.2.24 & Apache 2.0 \\
      \midrule
      \multirow{5}{*}{\textbf{Backend}} & fastapi & v0.121.2 & MIT \\
      & python-multipart & v0.0.20 & Apache 2.0 \\
      & uvicorn & v0.38.0 & BSD-3 \\
      & python-dotenv & v1.2.1 & BSD-3 \\
      & multiagent-firewall & v0.1.0 & MIT \\
      \midrule
      \textbf{Extension} & \multicolumn{3}{c}{No dependencies} \\
      \bottomrule
    \end{tabular}
  }
  }
\end{table}

\clearpage
\section{Test Environment and Reproducibility}
\label{sec:appendix-b}

\subsection{Hardware and Software Specifications}
All benchmarks reported in Section~\ref{sec:evaluation} were executed on a shared research infrastructure system with the following specifications:
\begin{itemize}[nosep]
  \item \textbf{CPU}: Dual Intel Xeon Platinum 8480C (x86\_64)
  \item \textbf{GPU}: NVIDIA H200
  \item \textbf{VRAM}: 139.81 GB (High Bandwidth Memory 3 Enhanced)
  \item \textbf{RAM}: 50 GB
  \item \textbf{Kernel}: Linux Kernel 6.8.0-79-generic
  \item \textbf{CUDA}: 12.8
  \item \textbf{Python}: 3.13.11
  \item \textbf{Local LLM Runtime}: vLLM 0.19.0 (\texttt{T11}: vLLM 0.20.1)
  \item \textbf{NER Model}: \texttt{gliner\_multi-v2.1}~\cite{gliner_2024}
  \item \textbf{LLM Detector Models}: \texttt{google/gemma-3-4b-it}~\cite{google_gemma_3_4b_it_2024}, \texttt{mistralai/Ministral-3-8B-Instruct-2512}~\cite{mistralai_ministral_3_8b_instruct_2512_2025}, \texttt{meta-llama/Llama-3.1-8B-Instruct}~\cite{meta_llama_3_1_8b_2024}, \texttt{google/gemma-4-E4B-it}~\cite{google_gemma_4_e4b_it_2026}
\end{itemize}

\subsection{Configuration}
The specific commit hash from where the tests were run is:
\vspace{0.2cm}
\begin{center}
  \texttt{6ca51f83d30ea16cbd4b508635e4aeeaa9782641}
\end{center}

The \texttt{LLM\_EXTRA\_PARAMS} were locked across all configurations as shown in Table~\ref{tab:llm-extra-params}.

\begin{table}[ht]
  \centering
  \caption{Environment configuration parameters for evaluation test execution.}
  \label{tab:environment-configuration-table}
  \resizebox{\textwidth}{!}{
  \renewcommand{\arraystretch}{0.85}
  \begin{tabular}{lll}
    \toprule
    \textbf{Parameter} & \textbf{Example Value} & \textbf{Description} \\
    \midrule
    \multicolumn{3}{c}{\textit{Dataset Configuration}} \\
    \texttt{INTEGRATION\_DATASET\_LOCALES} & \texttt{us} & Geographic locale for dataset. \\
    \texttt{INTEGRATION\_DATASET\_MAX\_CASES} & \texttt{500} & Number of test cases per run. \\
    \texttt{INTEGRATION\_DATASET\_SEED} & \texttt{100} & Random seed for reproducibility. \\
    \midrule
    \multicolumn{3}{c}{\textit{Analysis Configuration}} \\
    \texttt{CODE\_ANALYSIS\_ENABLED} & \texttt{false} & Enables code similarity detection. \\
    \midrule
    \multicolumn{3}{c}{\textit{LLM Configuration (Conditional execution only)}} \\
    \texttt{LLM\_PROVIDER} & \texttt{hosted\_vllm} & LLM service provider backend. \\
    \texttt{LLM\_MODEL} & \texttt{google/gemma-3-4b-it} & Specific LLM model identifier. \\
    \texttt{LLM\_EXTRA\_PARAMS} & \texttt{\{"temperature":0, ...\}} & JSON object with model parameters. \\
    \midrule
    \multicolumn{3}{c}{\textit{Detection Settings}} \\
    \texttt{NER\_ENABLED} & \texttt{false} & Toggles NER (GLiNER) detector. \\
    \texttt{FORCE\_LLM\_DETECTOR} & \texttt{false} & Controls LLM invocation mode. \\
    \texttt{MIN\_BLOCK\_LEVEL} & \texttt{low} & Minimum risk level to block. \\
    \bottomrule
  \end{tabular}
  }
\end{table}

\begin{table}[H]
  \centering
  \caption{Extra parameters defined for the LLM detection engine.}
  \label{tab:llm-extra-params}
  \resizebox{\textwidth}{!}{
  \renewcommand{\arraystretch}{0.85}
  \begin{tabular}{lll}
    \toprule
    \textbf{Parameter} & \textbf{Value} & \textbf{Description} \\
    \midrule
    temperature & \texttt{0} & Eliminates randomness. \\
    top\_p & \texttt{1.0} & Considers entire probability distribution. \\
    top\_k & \texttt{1} & Selects only the single most probable next word. \\
    frequency\_penalty & \texttt{0} & Prevents punishment for repeating words. \\
    presence\_penalty & \texttt{0} & Prevents topic drift based on prior appearance. \\
    repeat\_penalty & \texttt{1.0} & Neutral value for repetition. \\
    drop\_params & \texttt{true} & Strips unsupported metadata parameters. \\
    response\_format & \texttt{\{"type": "json\_object"\}} & Forces JSON response format. \\
    request\_timeout & \texttt{30} & Maximum wait time in seconds. \\
    max\_retries & \texttt{3} & Auto-retry attempts before returning error. \\
    \bottomrule
  \end{tabular}
  }
\end{table}

\end{document}